\documentclass[12pt,preprint,prd,nofootinbib]{revtex4}

\usepackage[pdftex]{graphicx}
\usepackage{latexsym,amsmath,amssymb,lmodern,float,url}
\usepackage{natbib}
\usepackage[pdftex,bookmarks,linktocpage,pdfpagelabels,plainpages=false,hyperfigures,linkcolor=blue,citecolor=blue]{hyperref} 
\usepackage{xspace}
\hypersetup{colorlinks=true}

\usepackage{graphicx}
\usepackage{amsmath}
\usepackage{amssymb}

\graphicspath{{./figs/}}

\newcommand\be{\begin{equation}}
\newcommand\ee{\end{equation}}
\newcommand\bea{\begin{eqnarray}}
\newcommand\eea{\end{eqnarray}}
\newcommand{\appropto}{\mathrel{\vcenter{
  \offinterlineskip\halign{\hfil$##$\cr
    \propto\cr\noalign{\kern2pt}\sim\cr\noalign{\kern-2pt}}}}}
\newcommand{\cns}{CE$\nu$NS\xspace}

\begin{document}
\preprint{MI-TH-1769}

\title{Accelerator and reactor complementarity\\ in  coherent neutrino scattering}
\author{James~B.~Dent$^{\bf a}$}
\author{Bhaskar~Dutta$^{\bf b}$}
\author{Shu~Liao$^{\bf b}$}
\author{Jayden~L.~Newstead$^{\bf c}$}
\author{Louis~E.~Strigari$^{\bf b}$}
\author{Joel~W.~Walker$^{\bf a}$}

\affiliation{$^{\bf a}$ Department of Physics, Sam Houston State University, Huntsville, TX 77341, USA}
\affiliation{$^{\bf b}$ Mitchell Institute for Fundamental Physics and Astronomy,
   Department of Physics and Astronomy, Texas A\&M University, College Station, TX 77845, USA}
\affiliation{$^{\bf c}$ Department of Physics, Arizona State University, Tempe, AZ 85287, USA}

\begin{abstract}
We study the complementarity between accelerator and reactor
coherent elastic neutrino-nucleus elastic scattering (CE$\nu$NS) 
experiments for constraining new physics in the form of non-standard neutrino interactions (NSI).
Firstly, considering just data from the recent observation by the COHERENT experiment,
we explore interpretive degeneracies that emerge when activating either two or four unknown NSI parameters.
Next, we demonstrate that simultaneous treatment of reactor and accelerator experiments,
each employing at least two distinct target materials, can break a degeneracy between up and down flavor-diagonal
NSI terms that survives analysis of neutrino oscillation experiments. Considering four flavor-diagonal
($ee/\mu\mu$) up and down-type NSI parameters, we find that all terms can be measured with
high local precision (to a width as small as $\sim$5\% in Fermi units)
by next-generation experiments, although discrete reflection ambiguities persist. 
\end{abstract}

\maketitle

\section{Introduction\label{sec:intro}}
\par The recent detection of coherent elastic neutrino-nucleus elastic scattering (CE$\nu$NS) by the
COHERENT experiment~\cite{Akimov:2017ade} has initiated a new chapter in the study of neutrino physics.
Using neutrinos from a stopped pion beam and a 14.6kg CsI[Na] target, COHERENT measured a best-fit
count of 134$\pm$22 CE$\nu$NS events, well in excess of the expected backgrounds. The measured rate
was found to be 77$\pm$16 percent of the Standard Model (SM) prediction, and constrains NSI for
some terms more strongly than previous deep inelastic scattering and oscillation
experiments~\cite{Akimov:2017ade,Coloma:2017ncl,Liao:2017uzy}.
Properties of the neutron distribution may also be extracted from the measurement~\cite{Cadeddu:2017etk}, and within 
certain frameworks models for light dark matter can also be probed~\cite{Ge:2017mcq}. 

\par The COHERENT CE$\nu$NS measurement in the stopped pion context may be complemented in the near future 
by currently commissioning reactor experiments~\cite{Aguilar-Arevalo:2016qen,Agnolet:2016zir,Billard:2016giu}, 
and also by dark matter experiments~\cite{Monroe:2007xp,Strigari:2009bq,Dutta:2017nht}. Reactor experiments are sensitive to electron-flavor NSI
(via $\bar{\nu}_e$), while COHERENT is sensitive to both muon and electron flavor NSI
(via $\pi^+ \rightarrow \nu_\mu\, [\mu^+ \rightarrow {\bar{\nu}}_\mu \nu_e e^+ ]$),
and dark matter experiments are sensitive to all flavor components from solar and astrophysical sources.
Due to the distinct average energies of the various neutrino sources, each of these experiments
probe the CE$\nu$NS cross section, and thus any prospective NSI, over different ranges of nuclear recoil energy. 

\par More generally, CE$\nu$NS experiments are complementary to neutrino oscillations probes of NSI,
because oscillation experiments are only sensitive to differences in flavor diagonal NSI components~\cite{Ohlsson:2012kf,Miranda:2015dra}
via differentially accrued phase. Further, neutrino oscillation experiments are unable to distinguish between up and down-type quarks,
even if only a single flavor NSI component in the mixing matrix is allowed to differ from zero.  Finally, the
forward scattering in matter-induced oscillation occurs at zero momentum transfer, such that even NSI mediated
by very light (relative to typical MeV nuclear scales) species will manifest as an effective Fermi-type point interaction.
Thus CE$\nu$NS experiments, and also deep inelastic scattering experiments~\cite{Coloma:2017egw}
(with much larger momentum transfer), fill important gaps in the study of NSI that are not possible to close 
using oscillation data alone.

\par Previous studies that have constrained NSI with both oscillation and scattering experiments typically
vary one or two NSI parameters when fitting to a given data set. This is justifiable, considering that oscillation
experiments are not able to distinguish between up/down type NSI, and specific experiments are most sensitive to
only a small number of NSI parameters. However, artificially disabling (or correlating) a number of NSI parameters
in order to collapse the parameter space also blinds the analysis to higher order symmetries between parameters
that may substantially weaken constraints derived from a particular experiment or set of experiments.

\par In this paper we study the prospects for breaking the up/down type NSI degeneracy in oscillation experiments
using a combination of COHERENT data with projected future data from reactor experiments. In contrast to previous studies,
we use the MultiNest algorithm to explore up to four NSI parameters simultaneously, and also marginalize over the uncertainties
on the neutrino fluxes from the sources and experimental backgrounds. Using realistic exposures for future reactor experiments,
we show that, combining all data sets, flavor diagonal terms can be measured to high precision. Constraints are strongest for low threshold
reactor measurements, which incorporate different detector targets.
Since we do not invoke the oscillation data directly, our results are strictly independent from and complementary
to constraints derived from experiments of that variety.

The outline of this paper is as follows.
Sec.~\ref{sec:NSI} briefly reviews  non-standard interactions and degeneracies of NSI coefficients, 
Sec.~\ref{sec:exps} discusses the experiments (current and future) from which we draw for constraints on NSI,
Sec.~\ref{sec:Bayes} introduces our multi-parameter Bayesian analysis framework,
Sec.~\ref{sec:results} discusses our results and we conclude in
Sec.~\ref{sec:conclusion}.

\section{Non-standard Interactions\label{sec:NSI}} 
\par The NSI landscape is vast, and it manifests several degeneracies that can obscure and ambiguate the interpretation
of experimental results.  In this section, we detail the definition of the CE$\nu$NS cross section, its dependence on NSI,
relevant degeneracies of the NSI parameter space, and the assumptions that we make for the various experimental setups. 

\subsection{Coherent Neutrino-Nucleus Scattering}
For mediator particles that are heavy compared to the typical momentum transfer $q$ of the CE$\nu$NS process, the NSI can be parameterized as
\begin{equation}\label{eq:nsi}
\mathcal{L}_{NSI}= -2\sqrt{2}G_{F}\sum_{\alpha,\beta,f}\bar{\nu}_{\alpha L}\gamma^{\mu}\nu_{\beta L}\left(\epsilon_{\alpha\beta}^{fL}\bar{f}_{L}\gamma_{\mu}f_{L}+\epsilon_{\alpha\beta}^{fR}\bar{f}_{R}\gamma_{\mu}f_{R}\right),
\end{equation}
where $\alpha$, $\beta = e, \mu, \tau$ indicate the neutrino flavor, $f$ the fermion type, and $L$/$R$ denote left and right-handed components.
Vector couplings are characterized by the spin-independent combination $\epsilon^{fV}_{\alpha\beta} = \epsilon^{fL}_{\alpha\beta}+\epsilon^{fR}_{\alpha\beta}$,
and axial-vector couplings by the orthogonal spin-dependent combination $\epsilon^{fA}_{\alpha\beta}=\epsilon^{fL}_{\alpha\beta}-\epsilon^{fR}_{\alpha\beta}$.
For the CE$\nu$NS process, the axial-vector contribution is negligible in comparison to the vector contribution (due to spin cancellation),
and will be neglected in the remainder of this work.  For mediators of mass $m_{X'}$ satisfying $m_{X'}^2 \lesssim q^2 \equiv 2 m_N E_R$ (where $m_N$
is the target mass, and $E_R$ its kinetic recoil energy), the Eq.~(\ref{eq:nsi}) NSI parameterization is altered by the onset 
of momentum dependence in the mediator's propagator. This creates a unique BSM signature in the recoil spectrum shape,
which turns up strongly as the energy $E_R$ decreases.

The differential cross-section for the CE$\nu$NS scattering process for an incident neutrino of energy $E_{\nu}$
can then be written as ~\cite{Barranco:2005yy}
\bea\label{eq:dcsnuclear}
\frac{d\sigma}{dE_R} = \frac{G_F^2Q_V^2}{2\pi}m_N\left(1-\left(\frac{m_NE_R}{E_{\nu}^2}\right) + \left(1-\frac{E_R}{E_{\nu}}\right)^2\right)F(q^2)
\eea
The function $F(q^2)$ is the nuclear form factor. It encodes the momentum dependence of the interaction, and is given by the Fourier transform
of the distribution of scattering sites in the nucleus. In this work we adopt the standard spin-independent Helm form factor \cite{Helm:1956zz}. 
The vector charge for a nucleus consisting of $Z$ protons and $N$ neutrons
incorporates both SM and NSI contributions, and is given by
\bea\nonumber\label{eq:vectorcharge}
Q_V^2 &\equiv& \left[Z(g_p^V + 2\epsilon_{\alpha\alpha}^{uV} + \epsilon_{\alpha\alpha}^d)+N(g_n^V + \epsilon_{\alpha\alpha}^{uV} + 2\epsilon_{\alpha\alpha}^d)\right]^2 \\&+& \sum_{\alpha\neq\beta}\left[Z(2\epsilon_{\alpha\beta}^{uV} + \epsilon_{\alpha\beta}^dV)+N(\epsilon_{\alpha\beta}^{uV} + 2\epsilon_{\alpha\beta}^{dV})\right]^2
\eea
The charges $g_V^p = 1/2 - 2\textrm{sin}^2\,\theta_W$ and $g_n^V = -1/2$ are the SM proton and neutron vector couplings, and $\theta_W$ is the weak
mixing angle. 


\subsection{Matter Induced Oscillation}
In the presence of NSI, neutrino oscillations in matter depend upon the effective NSI parameters
\bea
\epsilon_{\alpha\beta} = \sum_{f=u,d,e}Y_f(x)\epsilon_{\alpha\beta}^{f,V}
\label{eq:effective} 
\eea
where $Y_f(x)$ is the average of the $f/e$ density ratio. However, there are two key facts that make individual determination
of all parameters solely via oscillation experiments impossible: 1) oscillation is only sensitive to differences between 
diagonal NSI parameters, as $\epsilon_{\alpha\alpha}-\epsilon_{\beta\beta}$, and 2) transition probabilities
suffer from a generalized mass ordering degeneracy \cite{Coloma:2016gei}.  The generalized mass ordering degeneracy
(which arises as the LMA-Dark solution \cite{Miranda:2004nb} in the context of solar neutrinos) is due to an
invariance of the neutrino evolution with respect to the transformation $H_{vac}\rightarrow -H_{vac}^*$ of the vacuum Hamiltonian.
This identity, which is a manifestation of $CPT$ symmetry, is found to be functionally equivalent to a correlated
set of transformations of the neutrino mass-squared differences, mixing angles, and Dirac CP phase, accompanied by
a transformation of differences of the effective NSI parameters in the matter potential. Additionally, there is a
target-dependent degeneracy that arises due to the structure of the vector charge in Eq.~(\ref{eq:vectorcharge}).
Although results from oscillation experiments are not incorporated directly into the present analysis,
it is of interest to us to investigate the complementary manner in which CE$\nu$NS experiments
may constrain the effective linear combinations of parameters from Eq.~(\ref{eq:effective}) that are sensitive to the matter effect.
The nature of symmetries in the expression of $Q_V^2$, and the manner in which
CE$\nu$NS measurements can address those degeneracies, is explored in detail in the next subsection.



\subsection{Analysis of Degeneracies}
\label{sec:Degen}
We will seek now to identify possible modes of degeneracy in the
scattering rate for real-valued NSI coefficients with arbitrary sign.
For purpose of discussion, we will assume data that is fully consistent
with SM expectations, although no essential features are changed if this is not the case.
The first simple observation is that amplitudes with identical initial and final neutrino flavor
states will sum coherently, i.e. they will integrate over the constituent nuclear charges prior to squaring,
allowing for internal cancellation between couplings.  Conversely, distinct flavors will contribute
to the total scattering rate via $Q_V^2$ in Eq.~(\ref{eq:vectorcharge}) as a sum of squares.
In both cases, non-zero NSI terms may conspire in a manner that returns $Q_V^2$ to its SM value.

\begin{figure*}[ht]
\begin{tabular}{ccc}
\includegraphics[height=5cm]{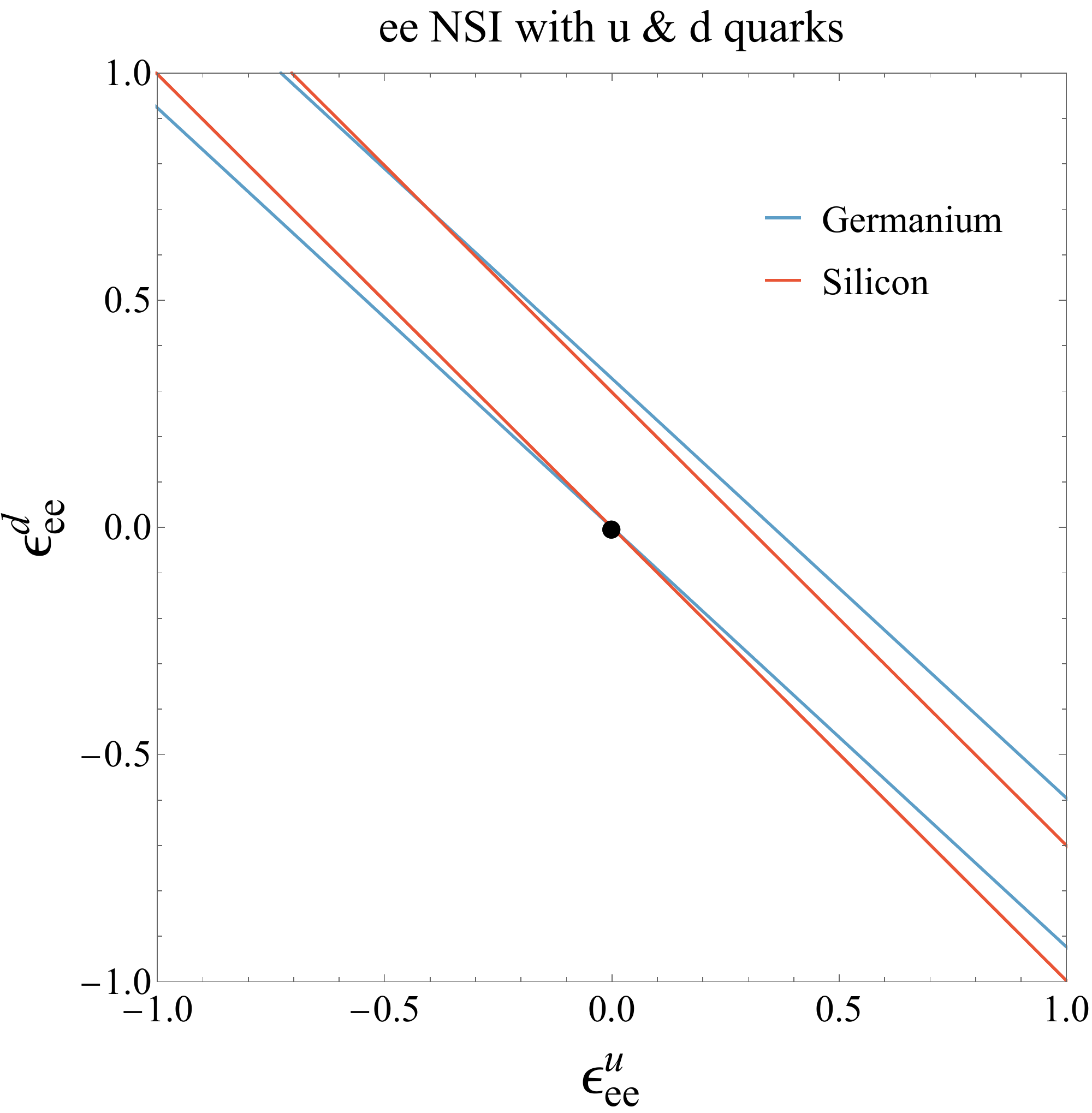} & 
\includegraphics[height=5cm]{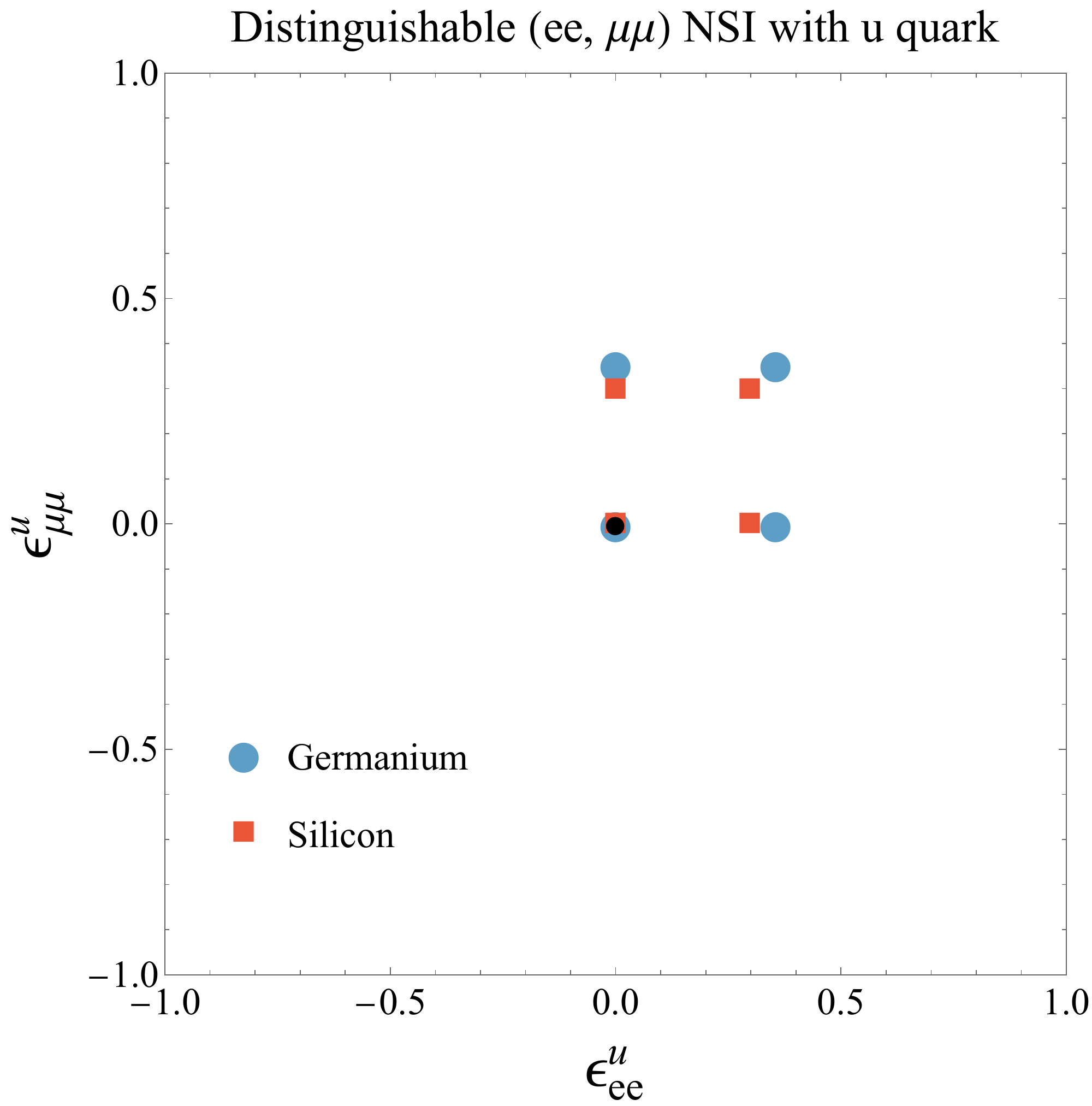} & 
\includegraphics[height=5cm]{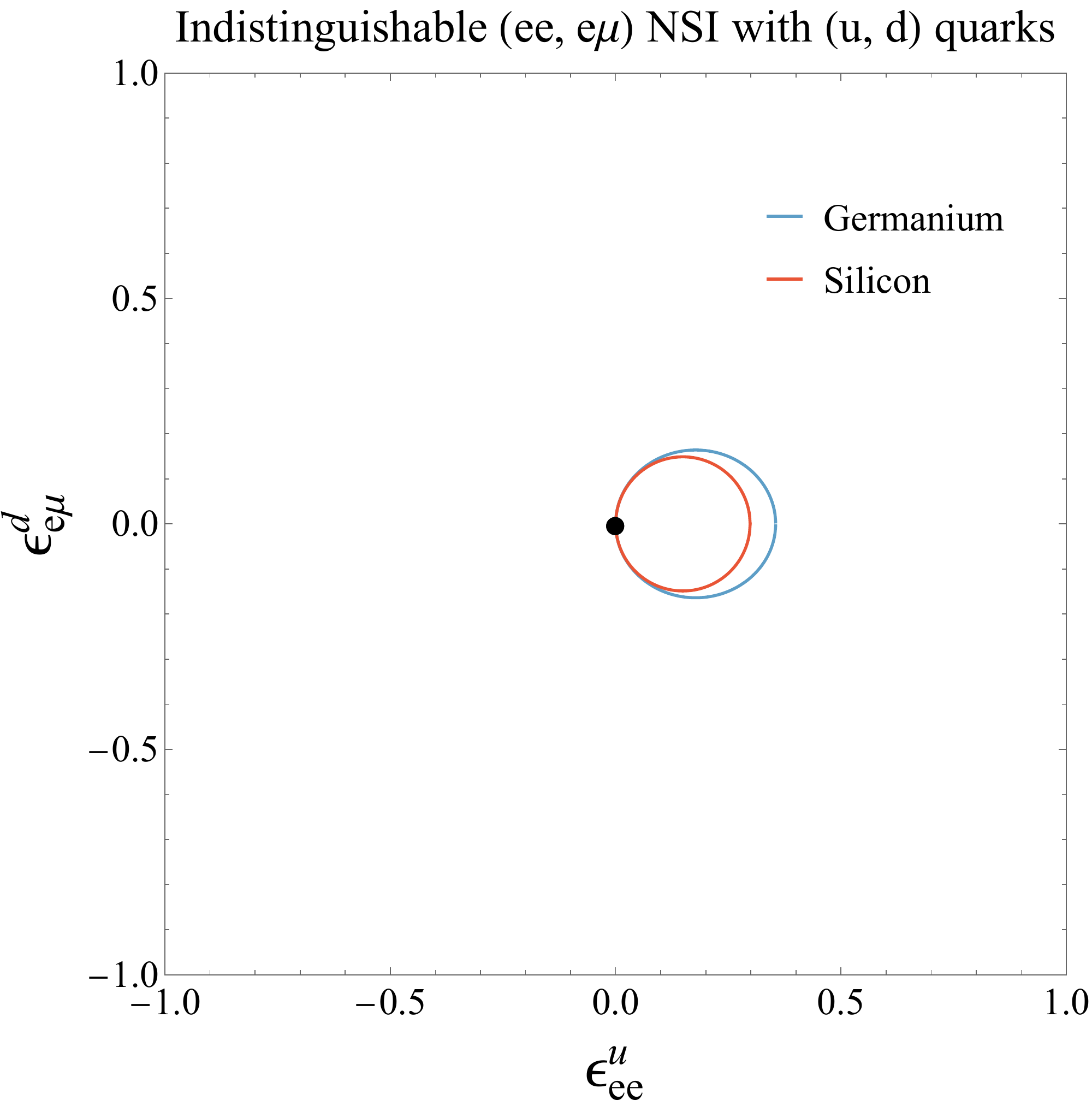} \\ 
\end{tabular}
\caption{The degeneracy-lifting effect of detector complementarity is exhibited for Si and Ge targets,
where NSI contributions interfere coherently within a single amplitude (left), or incoherently
as a sum of squares, where flavor contributions are either distinguishable (middle), or indistinguishable (right).}
\label{fig:contours}
\end{figure*}

In the former case, the solution for a constant scattering rate is linear,
forming an $(N-1)$-dimensional hyper-plane for $N$ free parameters.
For example, with spin-independent single-flavor NSI parameters
for the up and down quarks, there is a line of solutions
$\epsilon_{\alpha\alpha}^d = -(1+2\beta)/(2+\beta) \epsilon_{\alpha\alpha}^u$
that is indistinguishable from the standard model, with $\beta \equiv Z/N$.
There is additionally a discrete symmetry corresponding to inversion of the global sign inside a squared term.
This reflection is trivial for flavor-changing vertices with no SM coupling, but for
flavor-diagonal amplitudes it leads to a second parallel line of solutions that is disjoint from the SM, e.g.
$\epsilon_{\alpha\alpha}^d = -(1+2\beta)/(2+\beta) \epsilon_{\alpha\alpha}^u - 2(g_n+\beta g_p)/(1+2\beta)$.
If a dual observation is made via scattering off two different nuclei, then the linear degeneracies
are each (both the SM-connected and the SM-disjoint solutions) broken to a point of intersection by
variation of the slope.  For example, the intrinsic slopes $d \epsilon_{\alpha \alpha}^d/d \epsilon_{\alpha \alpha}^u$
of silicon and germanium targets are respectively $-0.997$ and $-0.923$ ($\beta = 0.992, 0.786$),
corresponding to an angular separation of $2.2$ degrees.
This scenario is depicted in the left-hand panel of Figure~\ref{fig:contours}.
Systematic and statistical uncertainties will obviously generate line broadening, such that the
residual consistency region is a pair of finite ``mirror image'' line segments.
Note for future reference that the angular separation between Ar and NaI is more shallow, approximately half of that for Si and Ge.

In the latter case, if the distinct final states are distinguishable, e.g. via timing information
in a single beam experiment, or via the combination of data from multiple experiments each
featuring a single known neutrino flavor, then the associated NSI constraints
will likewise be disconnected across flavors.
For example, with separable electron and muon vector NSI couplings to up quarks, a fixed point in the
$\epsilon_{ee}^u$ versus $\epsilon_{\mu\mu}^u$ plane will be preferred by data, along with three
images from sign-ambiguity under the pair of reflections
$\epsilon_{\alpha\alpha}^u \rightarrow - \epsilon_{\alpha\alpha}^u + 2(g_n+\beta g_p)/(1+2\beta)$.
This scenario is depicted in the center panel of Figure~\ref{fig:contours}.
Line broadening due to uncertainties will extend these points into solution patches of non-zero extent, and the advent of multiple
targets may further constrain any regions of overlap.
However, if only the aggregate rate is measurable, then a continuous generalization
of this discrete degeneracy emerges, which preserves the
sum of $M$ squared terms on the $(M-1)$-dimensional boundary of a hyper-ellipsoid.
For instance, with an electron-flavor $\nu_e$ neutrino source and vector NSI scattering
off up quarks into $\nu_e$ and $\nu_\mu$, there is a ring of solutions compatible with
the SM rate that is described by the equation
$[\epsilon_{e\mu}^u]^2 + [\epsilon_{ee}^u+(g_n+\beta g_p)/(1+2\beta)]^2 = [(g_n+\beta g_p)/(1+2\beta)]^2$.
Line broadening will lead in this case to a ring of non-zero width, separating under production to the
interior from overproduction to the exterior.
Application of different target nuclei will perturb the ring origin and radius,
conspiring to preserve an intersection at the SM solution point.
This overlap extends into a segment of arc after broadening.
Specifically, the origin will be translated along axes that carry a SM charge, and the
rings will be reflected symmetrically about the vector of displacement. 
If the NSI charges considered involve both up and down quarks, then an alternate target
will also affect the solution eccentricity, and multiple intersections are possible in the two-dimensional case.
This scenario is depicted in the right-hand panel of Figure~\ref{fig:contours}.

We will consider last the combination of these scenarios, with four non-zero NSI coefficients,
corresponding to vector flavor-diagonal scattering of electron and muon neutrinos
from up and down quarks (the scenario of primary interest to the current work).
If a single experiment is conducted, which cannot distinguish between flavor
(assuming comparable source fluxes), then any projection
into a two-dimensional space will exhibit an unconstrained continuum of solutions compatible with
the SM rate.  For example, any point in the $\epsilon_{ee}^u$ versus $\epsilon_{\mu\mu}^u$ plane
will be allowed, pending cancellation against a ring of counter-solutions in the hidden
$\epsilon_{ee}^d$ versus $\epsilon_{\mu\mu}^d$ plane.
One strategy for limiting this runaway is to impose a prior probability that penalizes
fine-tuning of NSI coefficients that are large relative to the weak force.
There are also several data-driven approaches to breaking the degeneracy, which we explore here.

The reactor-based \cns experiments have tremendous advantages in
neutrino flux (up to 5-6 orders of magnitude) over the COHERENT-styled stopped pion approach.  Even accounting for the higher neutrino
energies in the latter context (about 20-fold, corresponding to a relative scattering enhancement of
around 400), and the option to employ more simply-instrumented scintillating detectors of larger mass,
the advantage for future precision constraints on NSI coefficients is likely to tip toward the reactors.
The reactors are limited only to the testing of electron-sourced terms, but this turns to an advantage
in reducing ambiguity of NSI interpretations.  Confidence in measurement of the electron NSI at reactors
allows electron and muon effects to be separated at stopped pion sources, even without application of a timing cut. 
With this flavor separation, and with target complementarity at each experiment, the plane-filling
degeneracy is lifted to a residual solution of four SM-equivalent points
(corresponding to two pairs of sign ambiguities).
In practice, statistical and systematic effects enlarge the solution space.
The smaller target slope difference and lower flux for future stopped pion measurements 
at COHERENT can imply slower convergence for the muon constraints.  Imperfect
separation of flavor means that a residual ``weak coupling'' can persist
between between electron and muon NSI terms.

\section{Experiments\label{sec:exps}} 
\par In this section we discuss the experiments used to establish
current constraints on NSI parameters, and to project future constraints. 

\subsection{COHERENT} 
\par The COHERENT experiment uses a stopped pion beam which emits a prompt flux of $\nu_\mu$ from direct pion $\pi^+$
decay, and a delayed flux of $\bar{\nu}_\mu$ and $\nu_e$ from subsequent $\mu^+$ decay.
Though the recently published experimental results do not explicitly separate the flavors,
the prompt $\bar{\nu}_\mu$ may in principle be identified from a timing cut,
while the delayed components may be extracted from their spectral signatures.
To simulate the COHERENT measurement, we use the delivered protons on target combined with an
efficiency of 0.08 neutrinos per proton to obtain an average flux from all flavors at 20m of $1.05 \times 10^7$ cm$^{-2}$ s$^{-1}$.
Because the various flux components are not separated in the COHERENT analysis, we assign an uncertainty of 10\%
on each flux component, and marginalize over this uncertainty in the analysis below.
We further take an uncertainty of 5\% on the experimental background reported by COHERENT over
the relevant recoil energy range. We note that previous studies of NSI using the COHERENT
experiment~\cite{Scholberg:2005qs,Coloma:2017egw,Coloma:2017ncl,Liao:2017uzy} have instead assumed
that the per-flavor flux can be identified through a combination of timing and spectral signatures,
which would appear to be a reasonable future expectation. 

\par In addition to establishing constraints with current COHERENT data, we estimate future constraints from COHERENT data.
Motivated by a description of future experimental plans from the COHERENT collaboration,
we assume tonne-scale NaI and Ar targets, and larger exposures of 1 tonne-year.  To be conservative,
we retain the same uncertainties on the fluxes and the background, as described for the current COHERENT data.
In Table~\ref{tab:configurations} we show the details of our assumptions for each target. 


\subsection{Reactors} 
\par There have been several previous studies of NSI at nuclear reactors~\cite{Barranco:2005yy,Dutta:2015vwa,Dent:2016wcr,Dent:2016wor}.
The aforementioned reactor experiments operate at power ranges from $\sim$ MW-GW, and a range of distance from the reactor core from a
few meters to a few tens of meters.  To simulate a reactor experiment, we take a configuration that is broadly representative of
the entire class of experiments. Specifically, we adopt a baseline configuration of a 1~GW reactor with a detector site at $20$m
from the reactor core. Note that this configuration delivers a flux comparable to (or slightly larger than) that of a 1~MW
reactor at $\sim 1$m from the core. For either configuration, the $\bar{\nu}_e$ flux can be obtained via knowledge of the reactor
power along with the normalized antineutrino fission spectrum, which has been measured at various sites~\cite{An:2016srz}.
We consider Ge and Si detectors, and nuclear recoil thresholds that are attainable with present technology.
In Table~\ref{tab:configurations} we show the details of our assumptions for each target. 

\section{Multi-parameter Bayesian Framework\label{sec:Bayes}}
The parameter space of possible NSI is large, it is therefore historical practice to restrict one's attention to just
one or two such interactions at a time.  However, given certain previously elaborated degeneracies in the parameter space,
the constraints on allowed or disallowed regions will depend on which set of interactions are activated.
In this work we explore a set of NSI parameters within a Bayesian framework that can be easily extended
to include all NSI terms. Via Bayes' theorem, we calculate the posterior probability distribution,
$\mathcal{P}$, of the NSI parameter space, $\theta$, given some data $D$ and prior information, $I$:
\be
\mathcal{P}(\theta\vert D,I) = \frac{\mathcal{L}(D\vert\theta,I)\pi(\theta\vert I)}{\epsilon(D\vert I)}.
\ee
Here $\mathcal{L}(D\vert \theta, I)$ is the likelihood of a set of NSI parameters reproducing the observed
(or simulated) data. The prior probability, $\pi(\theta\vert I)$, is taken to be uniform for the NSI parameters
(i.e. there is no prior information), and taken to be Gaussian for the nuisance parameters.
The priors are summarized in Table~\ref{tabPriors}. Finally, the Bayesian evidence, $\epsilon(D\vert I)$,
serves as a normalization factor. Given that the NSI formalism under consideration does not change the shape of the
differential rate (this would change in the case of light mediators), we take the likelihood to be a product of Poisson probabilities for each experiment, as follows,
where $j$ runs over energy bins, and $i$ runs over the detectors used in a given experimental configuration.
\be
\mathcal{L}(D\vert\theta,I) = \prod_i \prod_j p(D_i\vert \lambda_i(\theta))
\ee
We will consider 4 configurations in this analysis, namely current data, future reactor data,
future accelerator data and a global analysis of both future reactor and accelerator data.
For the current configuration the data $D$ consists of the observed number of events at COHERENT (n=134). For our future projections, for accelerators we choose NaI and Ar detectors, and for reactors we choose Ge and Si for detectors. For all future configurations we take $D$ to be the Asimov (expected) dataset. The assumed exposures and thresholds for each target are shown in Table~\ref{tab:configurations}. To explore the parameter space we make use of the MultiNest package~\cite{Feroz:2008xx}, which implements the nested sampling algorithm due to Skilling~\cite{Skilling:2004}, and improved by Shaw~\cite{Shaw:2007jj}. This algorithm was developed for sampling the posteriors of high dimension parameter spaces which may contain multiple regions of high probability, as encountered in this analysis. The MultiNest sampling parameters were chosen as ($N_{\mathrm{live}}=1500$, $\mathrm{tol}=0.1$, $\mathrm{efr}=0.3$).
\begin{table}[ht]
\caption{Baseline priors used for the NSI parameters and nuisance parameters in this analysis.
Fluxes are per cm$^2\cdot$s, and backgrounds are per kg$\cdot$day$\cdot$keV.}
\centering
\label{tabPriors}
\begin{tabular}{|c|c|c|}
\hline
Parameter & Prior range & Scale \\
\hline
$\epsilon_{\alpha\alpha}^f$ & (-1.5, 1.5) & linear \\
SNS flux   & $(4.29\pm 0.43)\times 10^9$ & Gaussian \\
Reactor flux  & $(1.50\pm 0.03)\times 10^{12}$ & Gaussian \\
SNS background  & $(5\pm 0.25)\times 10^{-3}$ & Gaussian \\
Reactor background & $(1\pm 0.1)$ & Gaussian \\
\hline
\end{tabular}
\end{table}

\begin{table}[ht]
\caption{Experimental configurations used in this analysis}
\centering
\label{tab:configurations}
\begin{tabular}{|c|c|c|r|r|}
\hline
Name & Detector & Source & Exposure &Threshold \\
\hline
 Current (COHERENT)    & CsI & SNS (20m) & 4466 kg.days & 4.25 keV\\
\hline
 Future (reactor)      & Ge  & 1GW reactor (20m) & 10$^4$ kg.days & 100 eV \\
                       & Si  & 1GW reactor (20m) & 10$^4$ kg.days & 100 eV\\
\hline
 Future (accelerator)  & NaI & SNS (20m) & 1 tonne.year & 2 keV \\
                       & Ar  & SNS (20m) & 1 tonne.year & 30 keV \\
\hline
\end{tabular}
\end{table}

\section{Results} \label{sec:results}
\par
\begin{figure*}[ht]
\includegraphics[height=5.5cm]{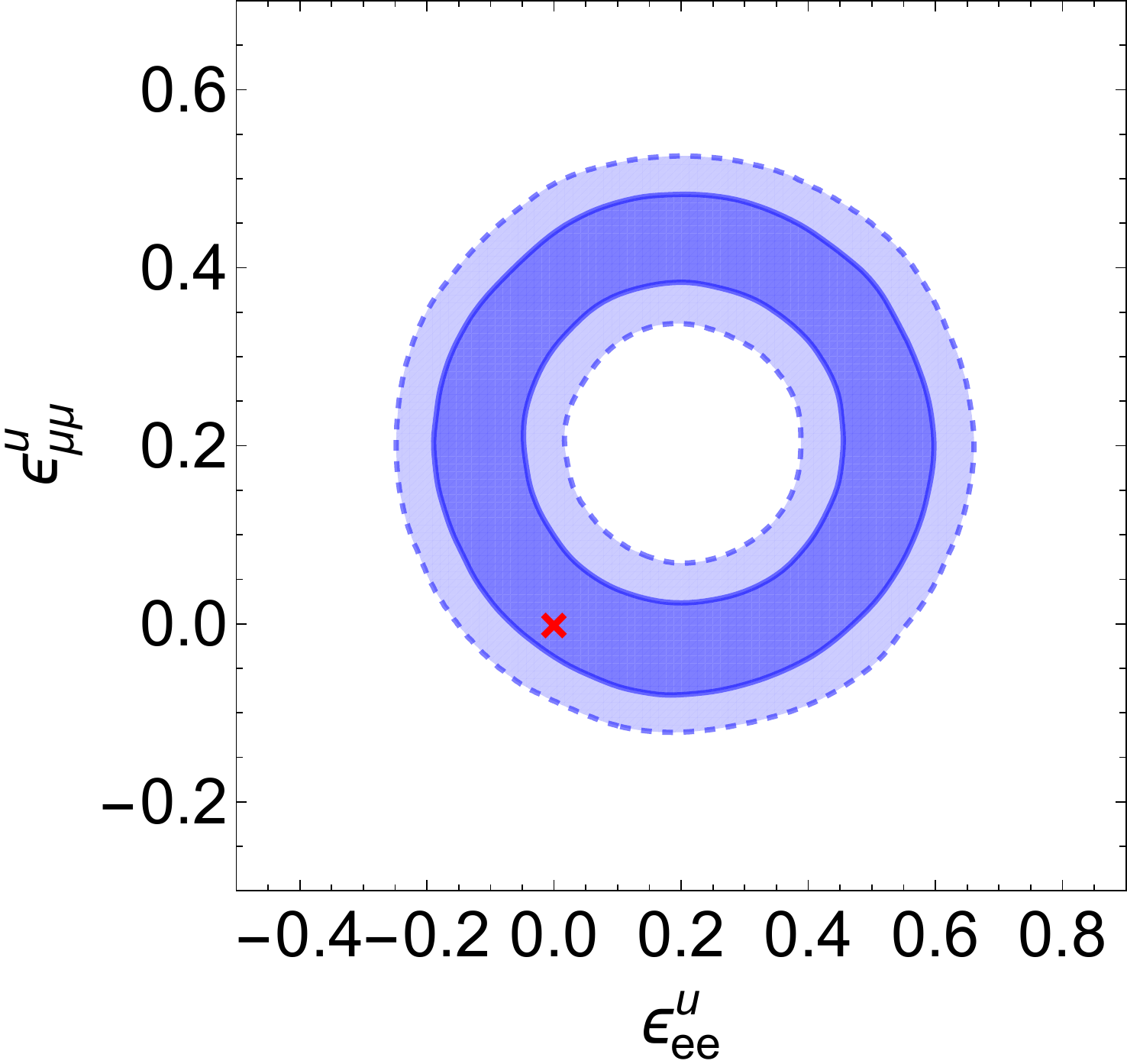}
\caption{Posterior probabilities of NSI parameters using the current COHERENT data, allowing for two non-zero
NSI parameters. The contours show the 68\% and 95\% credible regions, and the red cross indicates the Standard Model value.
}
\label{fig:current}
\end{figure*}

 We now apply the MultiNest formalism in order to constrain NSI parameter regions for different experimental configurations. 
We start by applying the formalism to current COHERENT data, allowing for two flavor diagonal $u$-type NSI to be non zero, with the result shown in Figure~\ref{fig:current}.
For this case, there is a ring of solutions in the $\epsilon_{\mu \mu}^u$ vs. $\epsilon_{e e}^u$ parameter space
because we are only considering flavor diagonal NSI, and the cross section is quadratically dependent on
NSI parameters. Similar solutions are seen for $d$-like NSI~\cite{Scholberg:2005qs,Coloma:2017ncl}.
Notice that the single experimental constraint has reduced a two-dimensional parameter space into a one-dimensional string of connected points.
However, if one allows for four NSI parameters to be non-zero simultaneously, no inferences can be made in the $\epsilon_{\mu \mu}^u$ vs. $\epsilon_{e e}^u$ parameter space,
owing to the possibility of cancellations between the up and down type NSI (see the full 4D posterior distribution for the current COHERENT data in the appendix for more detail).

\par
\begin{figure*}[ht]
\includegraphics[height=6.0cm]{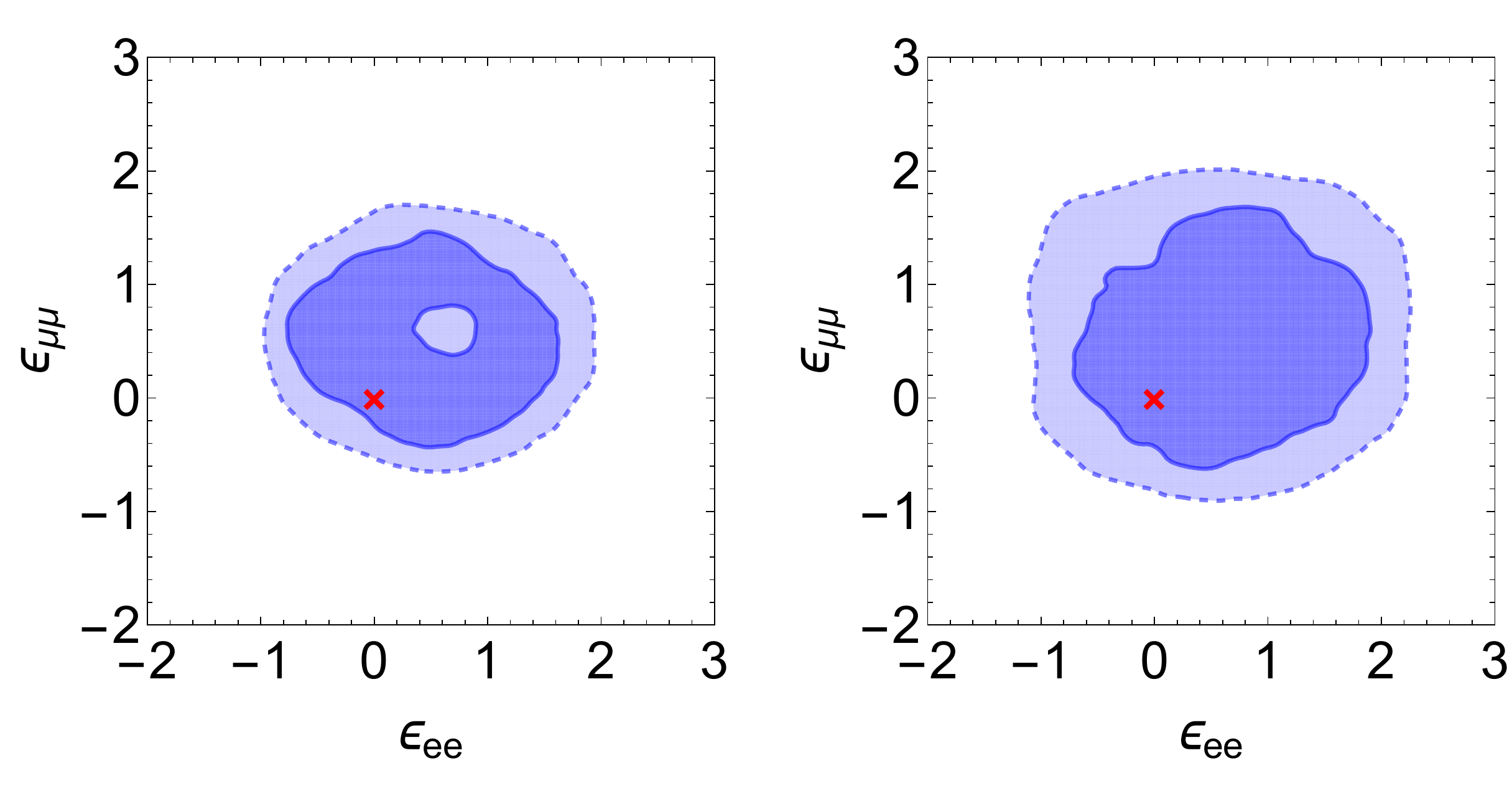} 
\caption{Posterior probabilities of effective NSI parameters using the current COHERENT data, allowing for four flavor diagonal parameters to be non-zero. The left panel takes flat priors on the $\epsilon$'s in the range $[-1.5:1.5]$, and the right panel uses flat priors in the range $[-2.5:2.5]$. The contours show the 68\% and 95\% credible regions, and the red crosses indicate the Standard Model value.  In particular, notice that these two-dimensional projections are space-filling with respect to definition of the prior, and thus do not represent any experimental constraints.
}
\label{fig:currenteffective}
\end{figure*}

 To examine the implications of current COHERENT data on the effective NSI parameters, as defined as in Eq.~(\ref{eq:effective}), we show the constraints in the $\epsilon_{ee}$ vs. $\epsilon_{\mu\mu}$ parameter space in
Figure~\ref{fig:currenteffective} when four NSI parameters are allowed
to be free. To examine the impact of deviation from our baseline priors
on the $\epsilon$'s, in this case we consider flat priors on each parameter in the 
range $[-1.5:1.5]$ and $[-2.5:2.5]$. Unlike in the two parameter case, we 
see that the constraints on effective NSI depend on the prior range for the 4 parameters. This is due to the cancellations between combinations of
$\epsilon^{u}_{\alpha\alpha}$ and $\epsilon^{d}_{\alpha\alpha}$
in Eq.~(\ref{eq:vectorcharge}), as discussed in Sec.~\ref{sec:Degen}, which makes the allowed parameter space larger as we increase the prior range for NSI parameters.
This is simply a reflection of the fact that current (one-dimensional) COHERENT data is insufficient isolate points, lines, or even surfaces from a four-dimensional parameter space. 

\par
\begin{figure*}[ht]
\centering
  \includegraphics[height=5.5cm]{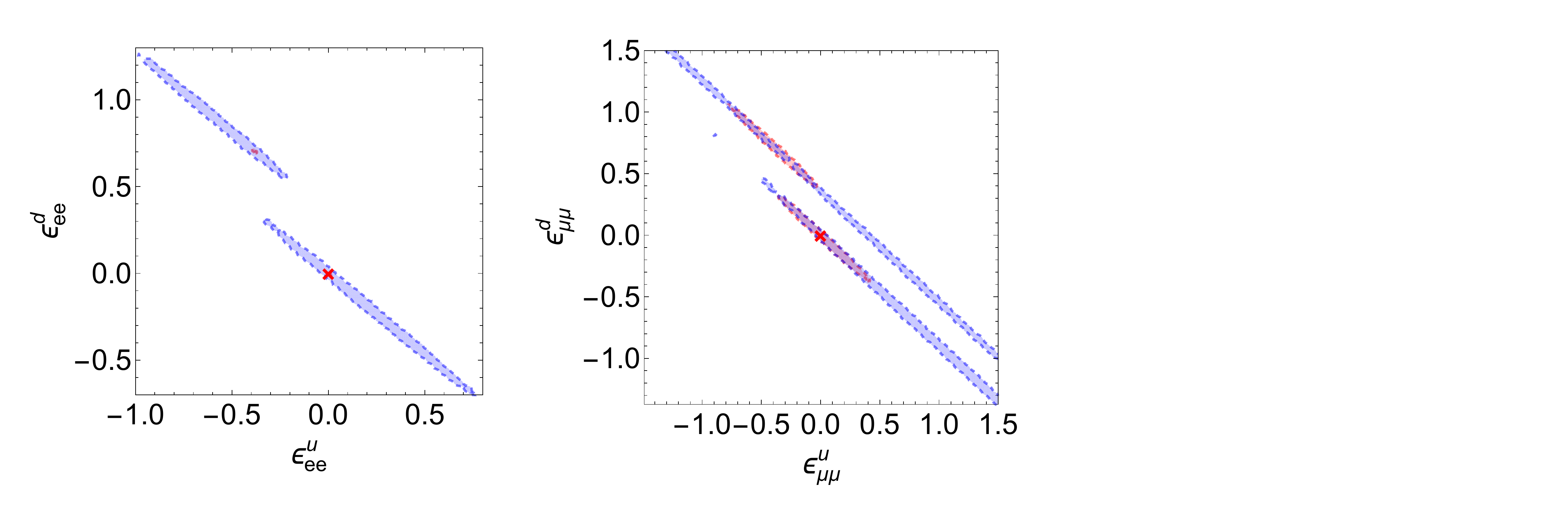}  
\caption{Comparison between unbinned simulation (blue) and binned simulation (red). For the unbinned case we take a single energy bin, while for the binned case we take ten energy bins. The contours show the 90\% credible regions and the red crosses indicate the simulated Standard Model value. 
}
\label{fig:1b10b}
\end{figure*}

 We now expand to consider the case of four free NSI parameters, with a simulated combination of future accelerator and reactor data.
With improved future data we anticipate better energy resolution, so we may consider the impact of binning the data in energy,
and compare to the results obtained to this point which have considered only a single energy bin. Shape information of the CE$\nu$NS
spectrum can provide information about whether the NSI is $\epsilon_{ee}$ or $\epsilon_{\mu\mu}$. This is due to the different energy
spectra of the neutrino species coming from the stopped pion source. As such, $\nu_e$ scattering events are more likely to produce lower energy
recoils, and $\nu_\mu$ and $\bar\nu_\mu$ are more likely to produce higher energy recoils. This difference allows for statistical discrimination
of the different flavors of NSI when spectral information is included in the likelihood. As an example, we compare a single bin reconstruction to
a ten bin reconstruction in Figure~\ref{fig:1b10b}. The extra shape information from CE$\nu$NS spectrum shrinks the credible regions in
parameter space. Motivated by this outcome, for the results in the remainder of this paper, we take the data to be distributed in ten energy bins.

\par
\begin{figure*}[ht]
\centering
  \includegraphics[height=5.5cm]{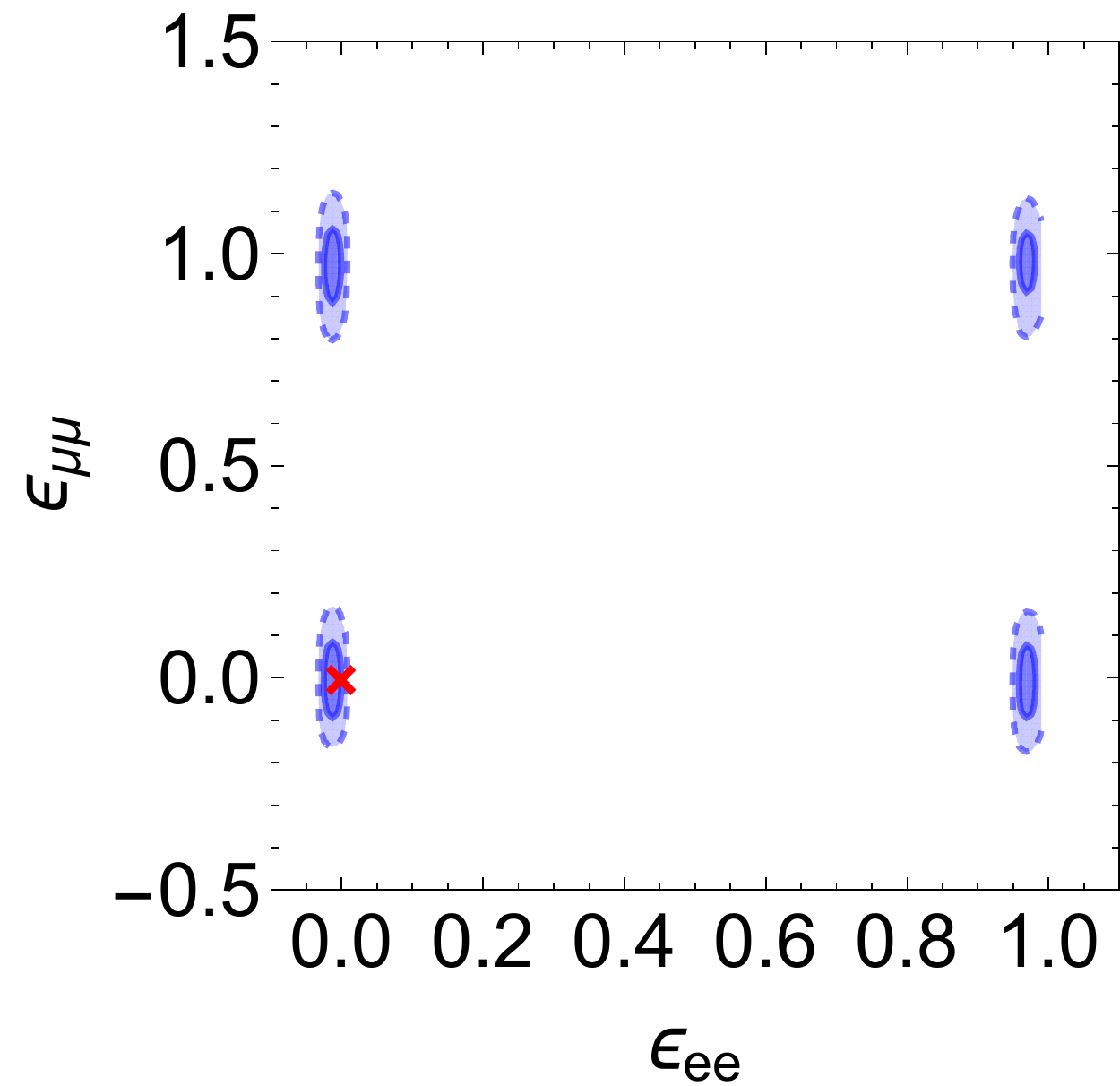} 
\caption{Projected posterior probabilities of effective NSI with future accelerator and reactor data, allowing for four flavor diagonal parameters to be non-zero. The contours show the 68\% and 95\% credible regions, and the red cross indicates the simulated Standard Model value.}
\label{fig:futureeffective}
\end{figure*}

\begin{figure*}[ht]
  \includegraphics[height=15cm]{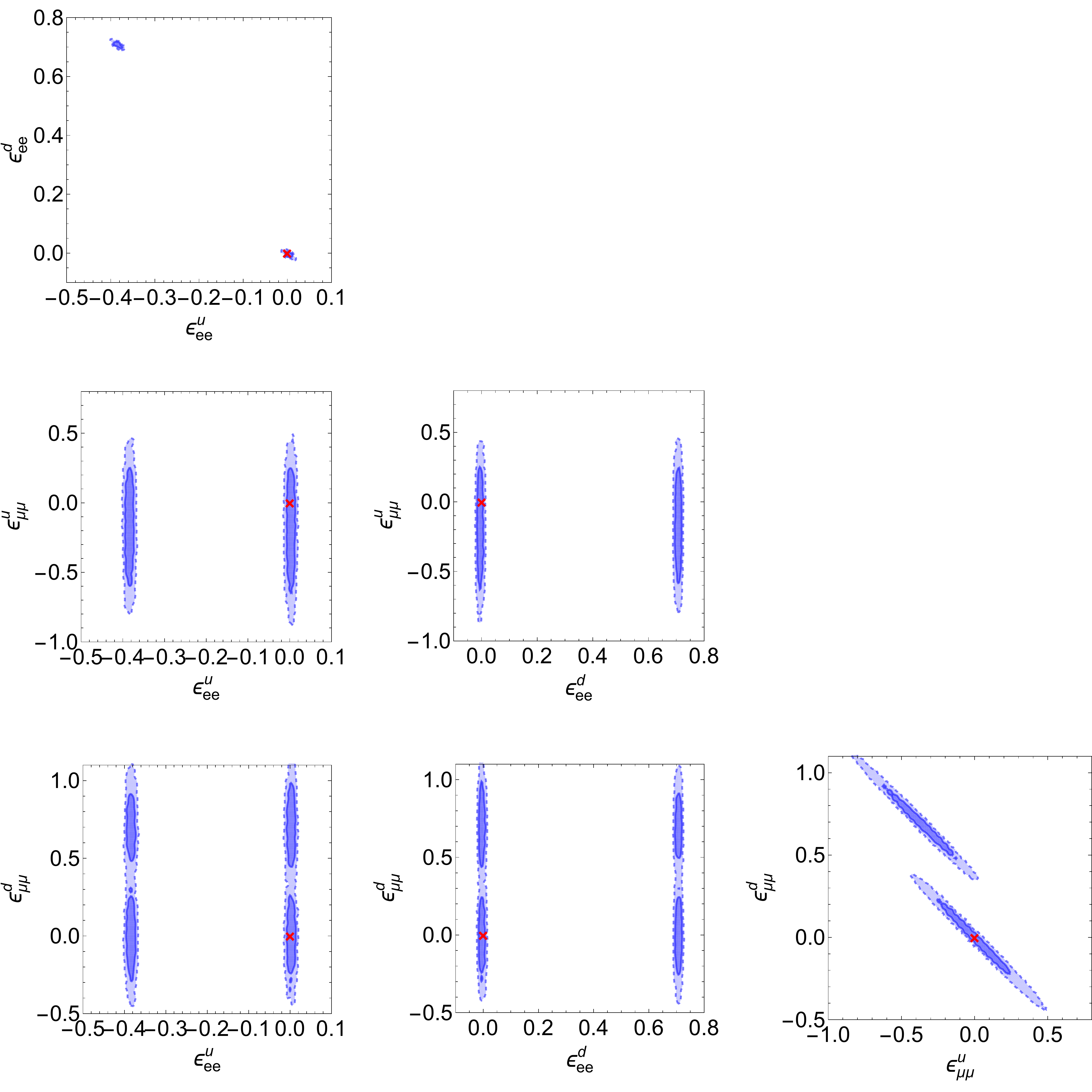}  
\caption{Projected posterior probabilities of the four NSI parameters with future accelerator and reactor data. Here we have marginalized over the uncertain experimental backgrounds and fluxes from the respective neutrino sources. The contours show the 68\% and 95\% credible regions, and the red cross indicates the simulated Standard Model value.}
\label{fig:future}
\end{figure*}

 In Figure~\ref{fig:futureeffective} we show the constraints on effective NSI, and in Figure~\ref{fig:future} we show the 
projected constraints on the individual NSI parameters. 
We note that relative to the two parameter case, when considering four free parameters, regions
of parameter space for large and negative $\epsilon$ are opened up. In addition, $d$-type NSI are
stronger constrained than $u$-type NSI. In comparison to the current COHERENT data, the space of degenerate solutions arising through combinations of
$\epsilon^{u}_{\alpha\alpha}$ and $\epsilon^{d}_{\alpha\alpha}$
is greatly reduced because the additional reactor data helps in breaking down the cancellations among these terms.

\par
\begin{figure*}[ht]
  \includegraphics[height=15cm]{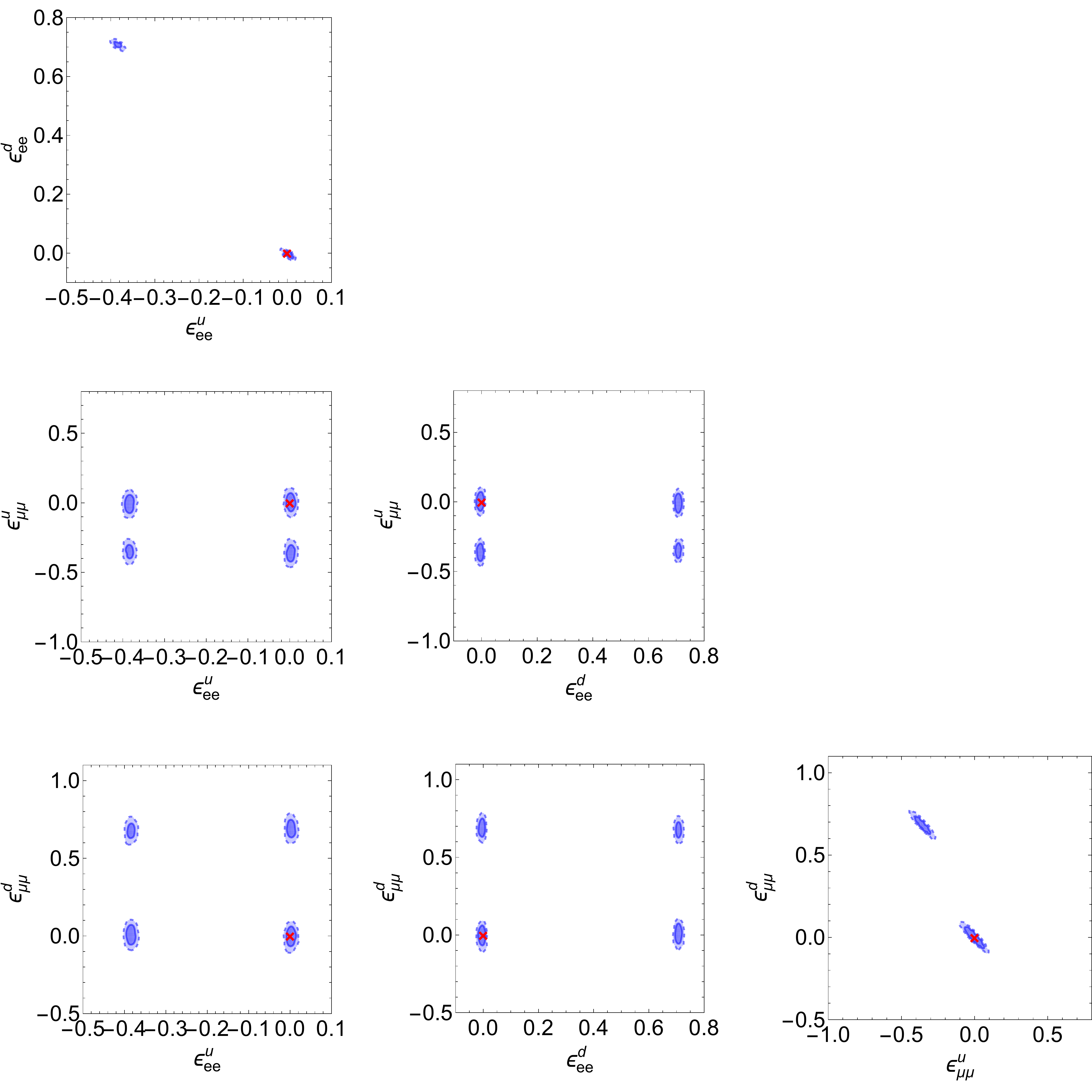}  \\
\caption{Projected posterior probabilities of the four NSI parameters with future accelerator and reactor data. Here we assume
zero experimental background for the accelerator detectors, all other uncertainties are marginalized over. The contours show the 68\% and 95\% credible regions, and the red cross indicates the simulated Standard Model value.}
\label{fig:noBackground}
\end{figure*}

 All of the above results include an experimental background and its associated uncertainty, and also include uncertainties on the neutrino fluxes from the sources. When considering the future accelerator data, the background and its uncertainty play a particularly important role in widening the allowed region in the NSI parameter space. This is demonstrated in Figure~\ref{fig:noBackground}, where we show the improvement that could be gained in the reconstruction if the experimental background was eliminated. In comparison with the contours in Figure~\ref{fig:future}, the allowed regions are much smaller.

\par Using the four parameter global fit to future reactor and accelerator experiments, we calculate the marginal $95\%$ credible intervals on the flavor diagonal terms and present them in Table~\ref{constraints}. We find that the lower uncertainty on the neutrino flux from reactors allows a very precise measurement of the $\epsilon_{ee}$ terms, up to the remaining ambiguities discussed in Sec.~\ref{sec:Degen}. The larger uncertainty of the neutrino flux from accelerators limits the precision of the $\epsilon_{\mu\mu}$ measurement. 

\begin{table}
  \caption{Projected $95\%$ credible intervals on each of the flavor diagonal parameters using future reactor and accelerator data\label{constraints}}
  \begin{tabular}{|c|c|c|c|}
  \hline 
  $\epsilon_{ee}^{u}$ & $\epsilon_{ee}^{d}$ & $\epsilon_{\mu\mu}^{u}$ & $\epsilon_{\mu\mu}^{d}$\tabularnewline
  \hline 
  $\left[-0.40,-0.37\right]\oplus\left[-0.01,0.02\right]$ & $\left[-0.02,0.02\right]\oplus\left[0.69,0.72\right]$ & $\left[-0.72,0.40\right]$ & $\left[-0.36,1.01\right]$\tabularnewline
  \hline 
  \end{tabular}
\end{table}

\section{Conclusion\label{sec:conclusion}}

The first measurement of the coherent elastic neutrino-nucleus scattering process by the COHERENT collaboration 
has ushered in a new era in the study of the neutrino sector, and has further demonstrated 
the fruitful avenue that neutrino physics continues to provide as a means of testing physics beyond the Standard Model.
The CE$\nu$NS process is already a powerful probe of non-standard neutrino interactions, and will soon extend its reach
with an array of new CE$\nu$NS reactor experiments slated to produce results in the near future.

In this work we have examined the complementarity provided by the COHERENT experiment's ability to study both muon-type
and electron-type NSI (due to their use of a stopped pion source) in an extremely low background environment, and that of upcoming 
nuclear reactor based experiments utilizing low-threshold cryogenic superconductor targets. A point of emphasis of the 
present work is on the ability of CE$\nu$NS measurements to begin to break various degeneracies that arise in
the NSI parameterization of neutrino interactions. 
For example, we have demonstrated the ability of combined analyses of reactor and stopped pion experiments with
multiple targets to probe the flavor diagonal up and down NSI parameter
degeneracy that arises in oscillation experiments.

Typically studies of the NSI parameter space have been carried out by examining the effects of (or constraints on)
either a single non-zero NSI parameter or a pair of non-zero NSI parameters.
 In this work we have utilized the MultiNest Bayesian 
inference tool in order to demonstrate the ability to constrain up to four NSI parameters with the current 
data from COHERENT, and have also provided future projections incorporating both additional COHERENT data and
reactor data generated by simulating a feasible near-term experiment. We find that all considered parameters can be measured
with high local precision (to a width as small as $\sim$5\% in Fermi units after marginalizing over other terms)
by next-generation experiments, although discrete reflection ambiguities persist.
However, this precision is most readily accessible with the high-flux reactor experiments, which are sensitive only to
the diagonal $\epsilon_{ee}$ terms, whereas similar measurements of the $\epsilon_{\mu\mu}$ coefficients will additionally
require very large exposures in future accelerator settings and greater improvements to the control of backgrounds.
It is apparent that existing and planned experimental programs designed to measure the CE$\nu$NS process
can act as precise probes of neutrino physics, providing unique insights into Standard Model physics and beyond.

\section{Acknowledgements}
BD and LES acknowledge support from DOE Grant de-sc0010813.
SL acknowledges support from NSF grant PHY-1522717.
JWW acknowledges support from NSF grant PHY-1521105.
JBD and JWW thank the Mitchell Institute for Fundamental Physics and Astronomy at 
Texas A\&M University for their generous hospitality, and JBD would like to thank the 
Mitchell Institute for support.

\appendix
\section{Current COHERENT 4D posterior distribution\label{app4D}}
\begin{figure*}[ht]
  \includegraphics[height=17cm]{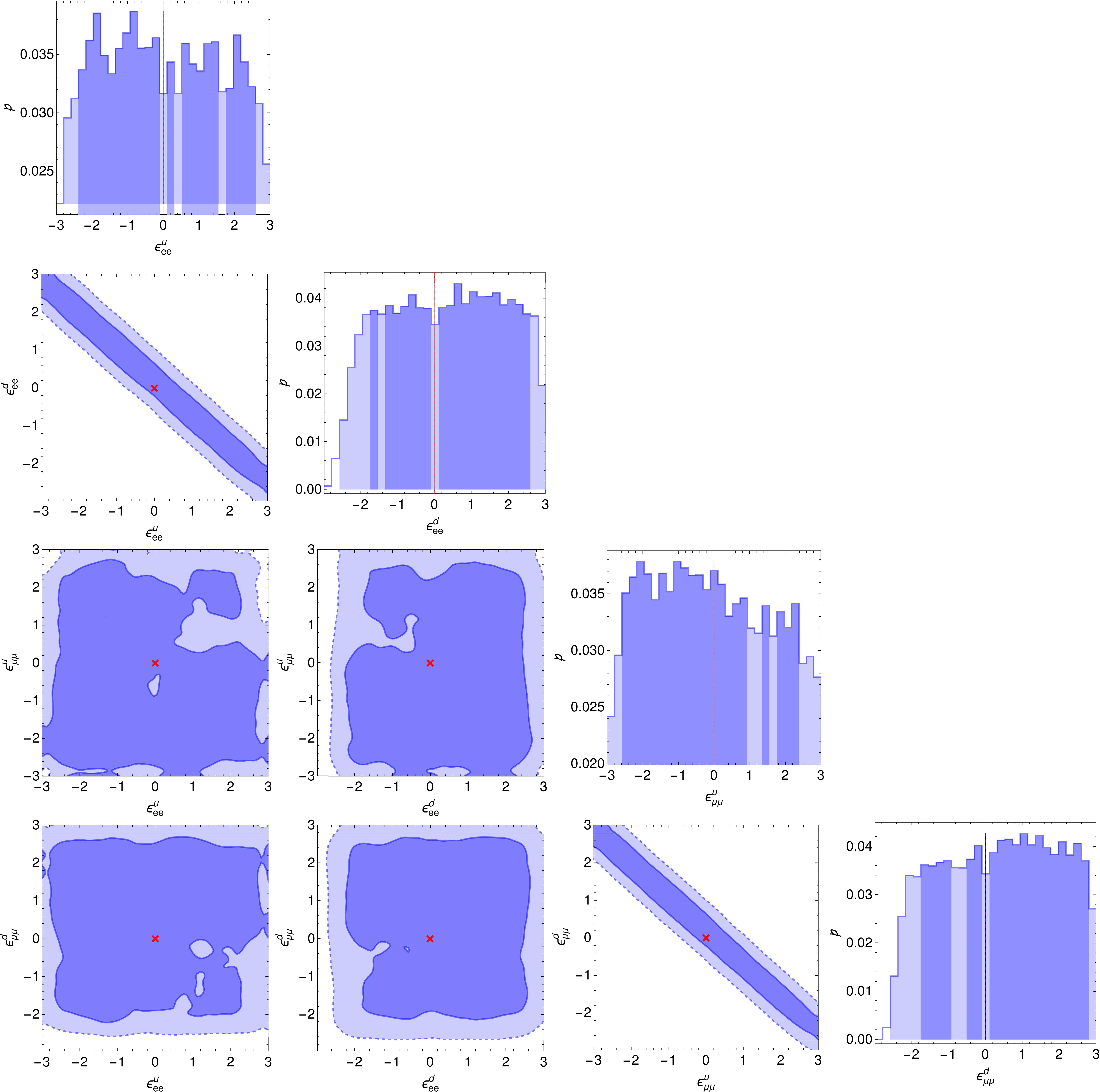}  
\caption{Current posterior distribution on the four NSI parameters with COHERENT data. Here flux and background uncertainties are marginalized over. The contours show the 68\% and 95\% credible regions.}
\label{fig:fullCOHERENT}
\end{figure*}

When one expands the NSI parameter space under consideration to include four flavor diagonal parameters cancellations can occur, opening up regions that may have been excluded when considering just two NSI parameters. This is illustrated with the posterior probability distribution of the 4D NSI parameter space for the current COHERENT data, shown in Fig.~\ref{fig:fullCOHERENT}. In comparison with the 2D NSI space of Fig.~\ref{fig:current} which produced a neat ring of solutions, now no inferences can be made in the equivalent 2D slice of parameter space. To highlight the degeneracies, here we extended beyond our baseline priors are take priors on the $\epsilon$'s in the range $[-3:3]$. 

\bibliography{main}

\end{document}